\documentclass[nopacs,twocolumn,preprintnumbers,notitlepage,amsmath,amssymb,superscriptaddress]{revtex4-2}

\usepackage[utf8]{inputenc}
\usepackage[utf8]{inputenc}
\usepackage{physics}
\usepackage{blindtext}
\usepackage{ dsfont }
\usepackage{tikz}
\usepackage{subfiles}
\usepackage{subcaption}
\usepackage[font=small,labelfont=bf,
   justification=justified,
   format=plain]{caption} 
\graphicspath{ {./Figures/} }
\usepackage{graphicx}
\usepackage{dcolumn}
\usepackage{bm}
\usepackage{amsmath}
\usepackage{amssymb}

\newcommand{\Sim}[1]{\mathrel{\mathop{\kern 0pt\sim}\limits_{#1}}}
\newcommand{\To}[1]{\mathrel{\mathop{\kern 0pt\to}\limits_{#1}}}
\newcommand{\Propto}[1]{\mathrel{\mathop{\kern 0pt\propto}\limits_{#1}}}
\makeatletter
\AtBeginDocument{\let\LS@rot\@undefined}
\makeatother
\usepackage[]{pdfpages}

\begin{document}

\title{Exact propagators of one-dimensional self-interacting random walks}

\author{Julien Br\'emont}
\affiliation{Laboratoire de Physique Th\'eorique de la Mati\`ere Condens\'ee, CNRS/Sorbonne Université, 
 4 Place Jussieu, 75005 Paris, France}
\affiliation{Laboratoire Jean Perrin, CNRS/Sorbonne Université, 
 4 Place Jussieu, 75005 Paris, France}
 \author{O. B\'enichou}
\affiliation{Laboratoire de Physique Th\'eorique de la Mati\`ere Condens\'ee, CNRS/Sorbonne Université, 
 4 Place Jussieu, 75005 Paris, France}
 \author{R. Voituriez}
\affiliation{Laboratoire de Physique Th\'eorique de la Mati\`ere Condens\'ee, CNRS/Sorbonne Université, 
 4 Place Jussieu, 75005 Paris, France}
\affiliation{Laboratoire Jean Perrin, CNRS/Sorbonne Université, 
 4 Place Jussieu, 75005 Paris, France}
\date{\today}
\begin{abstract}
    Self-interacting random walks (SIRWs) show long-range memory effects that result from the interaction of the random walker at time $t$ with the territory already visited at earlier times $t'<t$. This class of non-Markovian random walks has applications in contexts as diverse as foraging theory, the behaviour of living cells, and even machine learning.  Despite this importance and numerous theoretical efforts, the propagator, which is the distribution of the walker's position and arguably the most fundamental quantity to characterize the process, has so far remained out of reach for all but a single class of SIRW. Here we fill this gap and provide an exact and explicit expression for the propagator of two important classes of SIRWs, namely, the once-reinforced random walk and the polynomially self-repelling walk. These results give access to key observables, such as the diffusion coefficient, which  so far had not been determined. We also uncover an inherently non-Markovian mechanism that tends to drive the walker away from its starting point.
\end{abstract}

\maketitle

\begin{figure}
    \centering
    \includegraphics[width=\columnwidth]{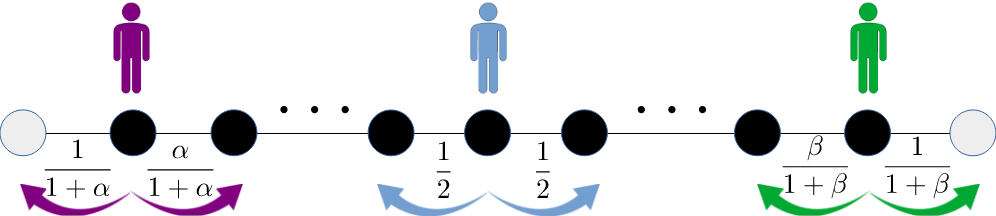}
    \caption{Sketch of a SATW. Unvisited sites are in gray, while already visited sites are in black. Different colors represent different situations : in the purple/green case, the RWer is at the left/right boundary of its span. In the blue case, it is in the bulk of its span. Jump probabilities are shown next to the arrows giving the  jump direction.}
    \label{fig:dessin}
\end{figure}
Consider the once-reinforced random walker (RWer) $(X_t)$ on $\mathbb{Z}$. This model, often called SATW (for self-attracting walk)  in the physics literature, is defined as follows. A nearest neighbour RWer starts at $X_0=0$. If it is on site $x$ at time $t$ and has not yet visited site $x+1$ (resp. $x-1$), it  jumps to this unvisited site at time $t+1$ with probability $1/(1+\beta)$ (resp. $1/(1+\alpha)$), while it  jumps to the visited site $x-1$ (resp. $x+1$) with probability $\beta/(1+\beta)$ (resp. $\alpha/(1+\alpha)$). If it has already visited both $x+1$ and $x-1$, it  jumps to either with equal probability. If $\alpha>1$ (or $\beta>1$), the RWer is thus attracted by the visited sites to its right (or left), whereas it is repelled  if $\alpha<1$ (or $\beta<1$); $\alpha=\beta=1$ restores the simple random walk (see FIG.\ref{fig:dessin})  \par 
This model belongs to the broad class of self-interacting random walks (SIRWs), which are characterized by long-lived  memory effects that emerge
from the interaction of the random walker at time $t$ with all the sites  that it has visited at earlier times $t' < t$. SIRWs, and in particular SATWs have clear applications
in various examples where a random walker induces non vanishing, local perturbations in its  environment—typically leaving  footprints along
its way  and, in return, is sensitive  to its own footprints,  being either attracted or repelled \cite{golestanian1,golestanian2}.
Such self-interactions have been reported qualitatively for ants or larger  animals that deposit
chemical cues as they move \cite{ants,animals} ; they also proved to be relevant to design sampling algorithms \cite{maggs}. Recently, they have  been identified for different types  of living cells \cite{cellattract,naturealex,lucas}, whose dynamics was shown to be quantitatively captured by the SATW model. It was found
in vitro in one- and two-dimensional geometries that these cells can chemically and mechanically modify  their local environment.
 In turn,  these non vanishing  footprints were shown to drastically modify the large scale cell dynamics, with
cells being effectively attracted by their footprints and thus preferentially remaining within the previously visited area. 
These experimental findings provide a prototypical example of attractive SATW, for which memory effects were
demonstrated  to have striking consequences on the dynamics of space exploration, such as aging (for $d=1,2$)  and subdiffusion (for $d=2$) \cite{cellattract,naturealex,lucas}.

\begin{figure}
    \centering
    \includegraphics[width=\columnwidth]{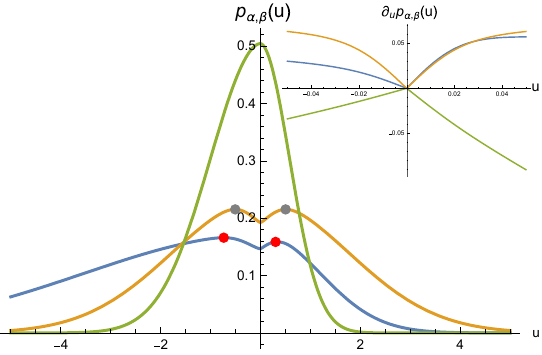}
    \caption{Scaling function of the propagator of the SATW \eqref{propag}. Different colors correspond to different values of the parameters $(\alpha,\beta)$. Blue corresponds to $(0.2,0.6)$, orange to $(0.4,0.4)$ and green to $(0.7,1.2)$. Colored dots show the local maxima in the distribution as predicted  by \eqref{approx-bump}. The inset shows the derivative $\partial_u p_{\alpha,\beta}(u)$ of the scaling function. Derivatives vanish at $u=0$.}
    \label{fig:pu}
\end{figure}

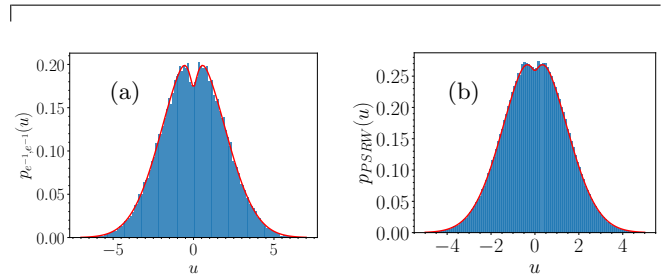
\begin{figure}
    \centering
    \begin{subfigure}[h]{0.48\columnwidth}
        \includegraphics[width=\textwidth]{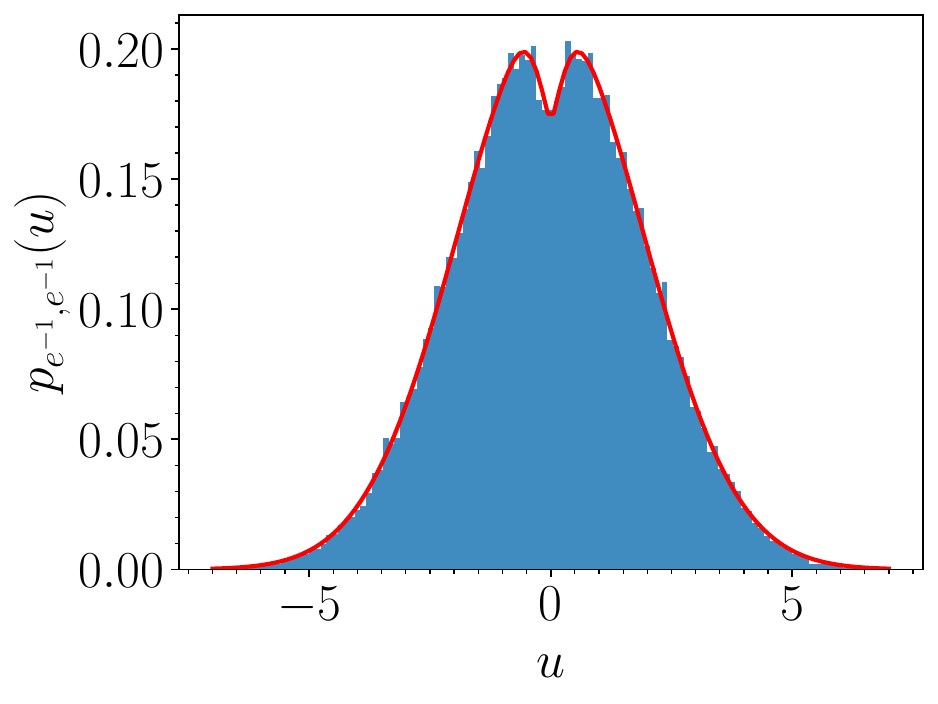}
        \begin{picture}(0,0)
    	\put(-20,80){(a)}
    	\end{picture}
    	\label{satw_num}
    	\vspace*{-20pt}
    \end{subfigure}
    \hfill
    \begin{subfigure}[h]{0.48\columnwidth}
        \includegraphics[width=\textwidth]{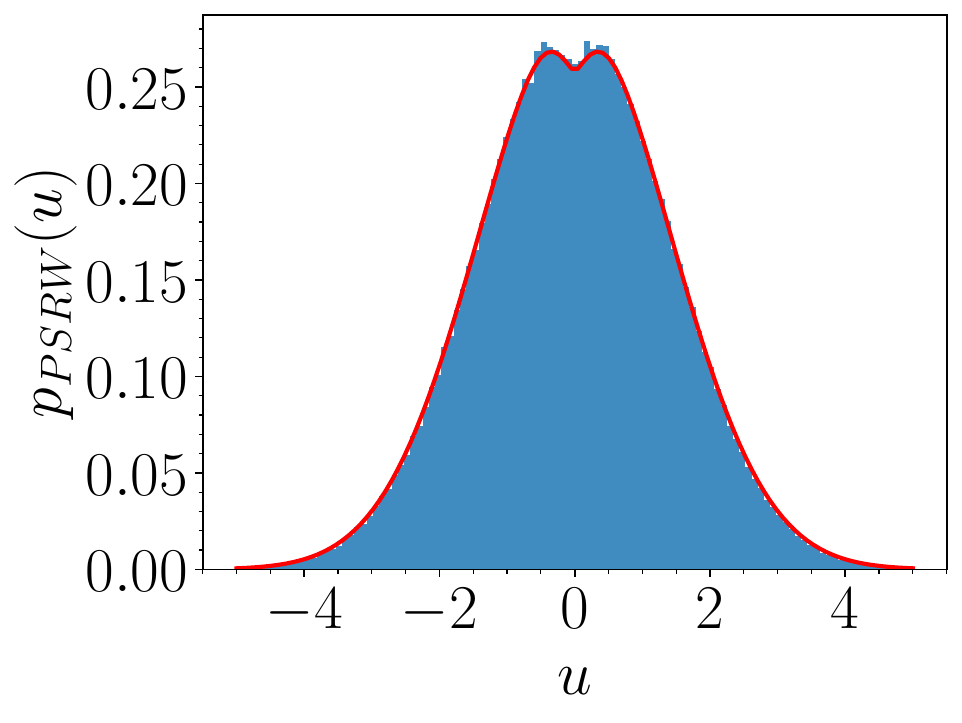}
        \begin{picture}(0,0)
		\put(-20,80){(b)}
		\end{picture}
		\label{psrw_num}
		\vspace*{-20pt}
    \end{subfigure}
    \caption{Propagators  of the  SATW and PSRW. The blue histograms show results of numerical simulations. (a) Scaling function of the propagator of a SATW with parameters $\alpha=\beta=1/e$. The red curve shows the theoretical result \eqref{propag}. Here, $u = x/\sqrt{2t}$. (b) Scaling function of the variable $u = \frac{x}{\sqrt{2t(2\gamma+1)}}$ for the polynomially self-repelling walk (PSRW), with exponent $\gamma=4$. The red curve shows $p_{1/2,1/2}(u)$ given by \eqref{propag}.}
    \label{fig:simus}
\end{figure}

On the theoretical side, SIRW models have attracted a lot of attention in both the mathematics \cite{pemantle,toth96,carmona-yor,diffusive-tsaw,dumaz} and physics \cite{parisi, prasad, sapo, alex-anomalous, alexprx, abc, structural, grassberger, naturealex, golestanian1,golestanian2} communities. Related examples of non-Markovian processes, in which the full history of past trajectories determines the future evolution, include self-avoiding walks \cite{hughes1995random}, locally activated RWs \cite{larw-pre,larw-new}  and RWs with reinforcement such as the elephant walks \cite{schutz,baur,bercu}. A wide variety of observables, such as first-passage properties \cite{doney,carmona-yor,alex-anomalous} or scaling exponents \cite{davis,toth96,sapo,prasad,diffusive-tsaw} have been calculated for SIRWs. Arguably, the propagator $P_{\alpha,\beta}(x,t)$, which is the probability for the RWer to be at site $x$ at time $t$, is of fundamental importance. However, of all the SIRW models, so far the propagator has  been determined explicitly only for the single example of the so-called "true self-avoiding walk" in \cite{dumaz}. This model  relies on the specific choice of self-interactions whose strength at each site depends linearly on the total number of visits up to time $t$ \cite{parisi} and is thus unbounded, in sharp contrast with the SATW. Although it has been known  \cite{davis} that the SATW is diffusive at long times, and that its propagator satisfies $P_{\alpha,\beta}(x,t) \sim p_{\alpha,\beta}\left(x/\sqrt{2t}\right)/\sqrt{2t}$ in the scaling limit $x, t\to \infty, x^2/t$ fixed, even the calculation of the diffusion coefficient has so far remained out of reach.  Progresses have been made, notably in \cite{toth96} and \cite{carmona-yor}, to determine the scaling function $p_{\alpha,\beta}$ for the SATW using results from generalised Ray-Knight theory, but these approaches did not provide explicit expressions for the propagator. In this Letter, using a different strategy, we obtain such an expression, and, as by-products, several other important quantities that so far remained unknown. We  first derive an exact expression of the joint distribution $\pi_{-m, \underline{k}}(t)$ of the minimum $-m$ and of the time $t$ needed to reach the maximum $k$ for the SATW. From this quantity, we explicitly obtain the joint distribution of the maximum, the minimum and the position of the RWer at time $t$. Taking the marginal of this distribution yields the following exact expression for the scaling function $p_{\alpha,\beta}$, which is our main result
\begin{widetext}
\begin{equation}
\begin{gathered}
\label{propag}
    p_{\alpha,\beta}\left(u = \frac{x}{\sqrt{2t}}\right) = \frac{(\beta-1)B\left(\beta ,\frac{\alpha +1}{2}\right)}{B(\alpha ,\beta ) \sqrt{\pi }} \sum_{n=0}^\infty  \frac{\left(\frac{1-\alpha}{2} \right)_n \left(\beta \right)_n}{ \left(\frac{1+2\beta+\alpha}{2} \right)_n} \frac{n+\frac{\beta}{2}}{(n+\frac{\beta-1}{2})(n+\frac{\beta+1}{2})} \frac{e^{-u^2(2n+\beta)^2}}{n!} \text{ for $u \geq 0$} 
\end{gathered}
\end{equation}
\end{widetext}
where $B(x,y)$ is the Beta function, and $(x)_n = x(x+1) \dots(x+n-1)$ is the Pochhammer symbol. The scaling function for $u<0$ follows from the space-reversal identity $P_{\alpha,\beta}(x,t) = P_{\beta,\alpha}(-x,t)$. Of note, as shown below, this function also provides the propagator of another class of SIRW, namely the polynomially self-repelling walk (PSRW) \cite{toth96}; it is displayed in FIG. \ref{fig:pu} and compared to  numerical simulations in FIG. \ref{fig:simus}. We provide below the main steps of the derivation of Eq.\eqref{propag}. \par 
\if We obtain several interesting quantities as by-products of our analysis.
First, we obtain a new, exact expression for the splitting probability of SATW, which we give in the SM.
Second, via a mapping to a run-and-tumble particle, we are able to solve an entirely new class of time-inhomogeneous telegrapher's equations. We show in SM that the normalised solution of the following equation
\begin{equation}
\label{telegrapher}
    \partial_L^2 P + \frac{2 \phi \mu}{\sinh L \mu} \partial_L P - \partial_c^2 P = 0, P(c,L=0) = \delta(c)
\end{equation}
is $P(c,L) = (2B(\frac{1}{2},\phi))^{-1} \partial_c B \left(\left(\frac{\sinh \left(\frac{\mu  c}{2}\right)}{\sinh \left(\frac{\mu  L}{2}\right)}\right)^2,\frac{1}{2},\phi \right)$ where $B(c,a,b) = \int_0^c t^{a-1}(1-t)^{b-1} dt$ is the incomplete Beta function. In the limit $\mu\to 0$, we recover a known telegrapher's equation \cite{poissondarboux}.
\fi

\textit{Joint distribution of the minimum and time to reach the maximum.}
Define $\pi_{-m, \underline{k}}(t)$ (resp. $\pi_{\underline{-m}, k}(t)$) as the probability that the trajectory of the RWer has minimum $-m$ (resp. maximum $k$) and that it reaches its maximum $k$ (resp. minimum $-m$) at time $t$. Note that we underline the last location reached by the RWer. In this section, as a first step to obtain the propagator, we derive an exact expression for this joint distribution, which  is interesting on its own as it  was shown to play an important role in trapping problems \cite{klinger}; in particular it gives access to classical splitting probabilities (see SM). \par 
Throughout, we denote the generating function of a given function $f(t)$ by $\hat{f}(\xi) = \sum_{t=0}^\infty \xi^t f(t)$. We will write $\xi = (1+s)^{-1}$, and $\mu = -\log((1-\sqrt{1-\xi^2})/\xi) \Sim{s\to 0} \sqrt{2s}$. A partition over the events where the RWer reaches either $k$ or $-m$ first yields
\begin{equation}
\label{func-eq}
    \hat{\pi}_{-m, \underline{k+1}} = \hat{\pi}_{-m, \underline{k}} \hat{F}^{\alpha,\beta}_{L}(\xi, +|+) + \hat{\pi}_{\underline{-m}, k} \hat{F}^{\alpha,\beta}_{L}(\xi, +|-)
\end{equation}
where $\hat{F}_L^{\alpha,\beta}(+|-)$ is the generating function of  $F_L^{\alpha,\beta}(\tau=t, \varepsilon|\varepsilon')$, which is  defined as follows. Denote  the span of the RWer at time $t$ (defined as the set of sites visited by the RWer up to time $t$) as $[-m,k]$, with $m,k\ge0$, and set  $L=k+m$. We denote by $\tau$ the time needed for the RWer starting from the boundary $\varepsilon = \pm$ of $[-m,k]$ ($+$ for $k$, $-$ for $-m$) to leave $[-m,k]$ through the boundary $\varepsilon'=\pm$ exactly at time $\tau$ ; $F_L^{\alpha,\beta}(\tau=t, \varepsilon|\varepsilon')$ is then defined as  the distribution of $\tau$. It is a $2\times 2$ matrix with indices $\pm$, which we call $F_L^{\alpha,\beta}(t)$ ; its generating function   
 $\hat{F}_L^{\alpha,\beta}$ is derived  in the SM with the help of a renewal equation \cite{hughes1995random}.
Equation \eqref{func-eq} (and its counterpart for $\hat{\pi}_{\underline{-m-1}, k}$) then gives access to the joint distribution  $\pi$ we are looking for. Making use of the following non trivial symmetry relation proved in SM
\begin{equation}
\label{ratio}
    \frac{\hat{\pi}_{-m, \underline{k}}}{\hat{\pi}_{\underline{-m}, k}} = \frac{\sinh(k \mu)}{\sinh(m \mu)},
\end{equation}
we obtain
\begin{equation}
\label{exact-pi}
    \hat{\pi}_{-m, \underline{k}} = \hat{\pi}_{-m, \underline{1}} \prod_{j=1}^{k-1} \left[\frac{\sinh m\mu}{\sinh j \mu} \hat{F}^{\alpha,\beta}_{m+j}(+|-) + \hat{F}^{\alpha,\beta}_{m+j}(+|+) \right]
\end{equation}
where $\hat{\pi}_{-m, \underline{1}} = \hat{F}_m^{\alpha,\beta}(+|-) \prod_{j=0}^{m-1} \hat{F}_j^{\alpha,\beta}(-|-)$, which holds for all values of parameters. In particular, in the scaling  regime $L\to \infty, s\to 0, L\sqrt{s}$ fixed, we find 
\begin{equation}
    \label{fpt-span-scaling}
    \hat{F}_L^{\alpha,\beta} \sim \mathds{1}_2 + \sqrt{2s} \begin{bmatrix} -\beta \coth(L\sqrt{2s}) & \frac{\alpha}{\sinh L\sqrt{2s}} \\ \frac{\beta}{\sinh L\sqrt{2s}} & -\alpha \coth(L\sqrt{2s})
    \end{bmatrix}
\end{equation}
where $\mathds{1}_2$ is the identity matrix. Taking the scaling limit of \eqref{exact-pi} with the help of \eqref{fpt-span-scaling}, in the regime $k,m \gg 1, s\to 0, L\sqrt{s}$ fixed, we finally obtain (see SM):
\begin{equation}
\label{pi-scaling}
    \hat{\pi}_{-m, \underline{k}} \sim \frac{\sqrt{2s}}{B(\alpha,\beta)} \frac{\sinh^{\beta-1} (m\sqrt{2s}) \sinh^\alpha(k\sqrt{2s})}{\sinh^{\alpha+\beta}((k+m)\sqrt{2s})}.
\end{equation}

\textit{Joint distribution of minimum, maximum, and position.}
Next, we derive the scaling limit  of the  generating function  of the joint distribution $Q(k,m,x,t)$ of the minimum, maximum, and position of the RWer. A partition over the events where  the most recent visited site is either  $k$ or $-m$ yields
\begin{equation}
\label{exact-joint-maxmin-pos}
    \hat{Q}(k,m,x,\xi) = \hat{\pi}_{\underline{k},-m} \hat{P}_s^{m,k}(x|+) + \hat{\pi}_{\underline{-m},k} \hat{P}_s^{m,k}(x|-)
\end{equation}
where $\hat{P}_s^{m,k}(x|\varepsilon)$ is the generating function of $P_s^{m,k}(x,\tau|\varepsilon)$ defined as 
the probability that the RWer is on site $x$ at time $t+\tau$ and has not left the span  $[-m,k]$, knowing that it was at  the boundary $\varepsilon = \pm$ of the span at time $t$. Using a renewal approach \cite{hughes1995random}, we compute this quantity  in SM and show that, in the scaling limit $m, k \gg 1, s \to 0, L\sqrt{s}$ fixed, one has 
\begin{equation}
    \hat{P}_s^{m,k}(x|-) \sim \frac{2\alpha (\sinh(k-x)\sqrt{2s})}{\sinh L\sqrt{2s}}.
\end{equation}
Finally, we have obtained all terms in the rhs of \eqref{exact-joint-maxmin-pos}, which provides an exact determination of $ \hat{Q}(k,m,x,s)$ for all values of the parameters. In the asymptotic limit $k,m \gg 1, s \to 0, L\sqrt{s}$ fixed, this gives the simpler form
\begin{widetext}
\begin{equation}
\label{jointdist-min-max}
    \hat{Q}(k,m,x,s) \sim \frac{2\sqrt{2s}}{B(\alpha,\beta)} \frac{\sinh^\beta (m\sqrt{2s}) \sinh^\alpha(k\sqrt{2s})}{\sinh^{\alpha+\beta+1}((k+m)\sqrt{2s})} \left( \beta\frac{ \sinh(m+x)\sqrt{2s}}{\sinh(m\sqrt{2s})} + \alpha\frac{ \sinh(k-x)\sqrt{2s}}{\sinh(k\sqrt{2s})}\right).
\end{equation}
\end{widetext}
The joint distribution  $Q(k,m,x,t)$ readily gives access  to the propagator with absorbing boundaries  $\tilde{P}_{a,b}(x,t)$, which is the probability that the RWer, starting at $0$ at time $t=0$, is at $x$ at time $t$, knowing that it is absorbed if it hits the sites $-a,b$. For $x\geq 0$, this can be written as $\tilde{P}_{a,b}(x,t) = \sum_{k=x}^{b-1} \sum_{m=0}^{a-1} Q(k,m,x,t)$. In the scaling limit, the sums can be written as integrals, which we compute explicitly  for $a,b \to \infty$ (see SM) : this limit yields the unconstrained propagator \eqref{propag}, which is the main result of this Letter. Besides its intrinsic importance, this result reveals   several remarkable features of the SATW. \par
\begin{figure}
    \centering
    \begin{subfigure}[h]{0.54\columnwidth}
        \includegraphics[width=\textwidth]{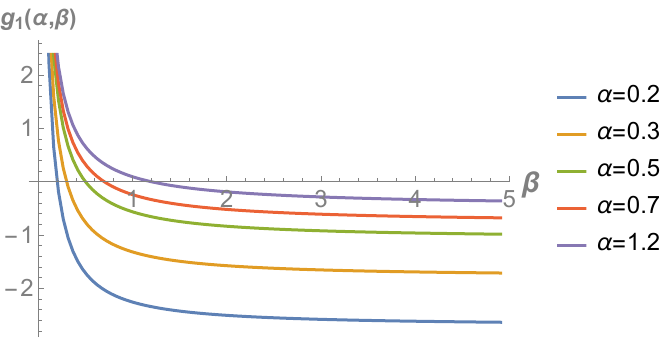}
        \begin{picture}(0,0)
    	\put(-20,50){(a)}
    	\end{picture}
    	\label{satw_num}
    	\vspace*{-20pt}
    \end{subfigure}
    \hfill
    \begin{subfigure}[h]{0.42\columnwidth}
        \includegraphics[width=\textwidth]{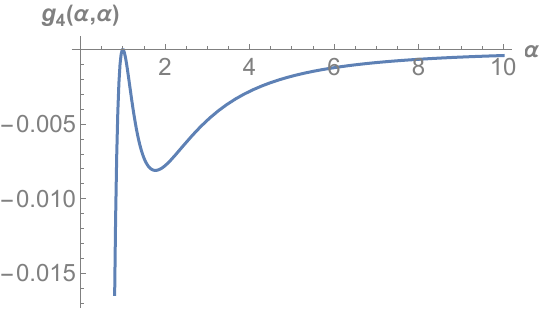}
        \begin{picture}(0,0)
		\put(-10,30){(b)}
		\end{picture}
		\label{psrw_num}
		\vspace*{-20pt}
    \end{subfigure}
    \caption{(a) Scaled mean $g_1(\alpha,\beta) \sim \frac{\langle X_t \rangle}{\sqrt{2t}}$ plotted for different values of $\alpha$. As expected, it becomes negative for $\beta > \alpha$, i.e. when the RWer is more likely to increment its span to the left than to the right. (b) Scaled fourth cumulant $g_4(\alpha,\alpha) \sim \frac{\kappa_4(t)}{3t ^2}$ for symmetric SATW $\alpha=\beta$. Note that it vanishes only for $\alpha=1$.}
    \label{fig:cumulants}
\end{figure}
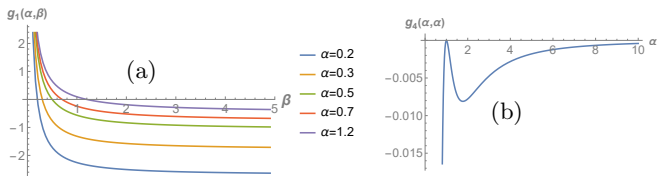

 
\textit{Moments of the position $X_t$. } First, Eq. \eqref{propag}  gives access to the moments $\langle X_t^n \rangle$ of the position of the RWer at time $t$.  In the general case $\alpha \neq \beta$, the mean position $\langle X_t \rangle$ is non-zero ; it is obtained exactly from  \eqref{propag} and can be written $\langle X_t \rangle = g_1(\alpha,\beta) \sqrt{2t}$ where $g_1(\alpha,\beta) = -g_1(\beta,\alpha)$ is explicitly given in the SM and plotted in  FIG. \ref{fig:cumulants}(a). Note the $\sqrt{t}$ scaling of the mean position (expected from the general scaling form of $p_{\alpha,\beta}$), which is at odds with the classical case of diffusion with constant drift.  In limit cases, simple forms of $ g_1(\alpha,\beta)$ can be derived : for $\alpha=1$, one has  $g_1(1,\beta) = \left(1-\beta\right)/(\sqrt{\pi} \beta)$, and  we find the asymptotic behaviour 
$g_1(\alpha,\beta) \Sim{\alpha \to 0} -1/(\alpha \sqrt{\pi})$. \par 
Next, the variance $V(X_t) = \langle X^2_t \rangle - \langle X_t \rangle^2\Sim{t \to \infty} 2D_{\alpha,\beta} t$ is also obtained from Eq. \eqref{propag} and defines the diffusion coefficient $D_{\alpha,\beta}$, which  is a symmetric function of $\alpha,\beta$ given in SM.  Limiting cases yield the simple expressions  $D_{1,\beta} = \left(\pi  \beta  (\beta -1)-2(\beta -1)^2+\pi \right)/(2\pi  \beta ^2)$ and
$D_{\alpha,\beta} \Sim{\alpha \to 0}  (\pi-2)/(2\pi \alpha^2)$. Remarkably,  our approach provides in particular an explicit expression, so far unknown,  in the symmetric case $\alpha=\beta$ :
\begin{widetext}
\begin{equation}\label{D}
    D_{\alpha,\alpha} = \frac{1}{2} + \frac{(1-\alpha ) B\left(\frac{\alpha +1}{2},\alpha \right)}{B(\alpha ,\alpha )} \frac{ \, _4F_3\left(\frac{1-\alpha}{2},\frac{\alpha }{2},\frac{\alpha }{2},\alpha ;\frac{\alpha }{2}+1,\frac{\alpha }{2}+1,\frac{3 \alpha }{2}+\frac{1}{2};1\right)}{\alpha ^2}
\end{equation} 
\end{widetext}
where $_pF_q$ is a hypergeometric function. 

Finally, Eq. \eqref{propag} shows that the propagator is not Gaussian (for $(\alpha,\beta)\not=(1,1)$), even if it has  a Gaussian tail $p_{\alpha,\beta}(u) \Propto{u\to \infty} e^{-u^2 \beta ^2}$. Of note, if $\alpha = 2n+1$ is an odd integer, the Pochhammer symbol $((1-\alpha)/2)_k$ is zero for $k > n$, so that   $p_{\alpha,\beta}$ is a \textit{finite} sum of $n+1$ Gaussian functions.   Non-Gaussianity can be quantified by  the fourth cumulant $\kappa_4(t) = \langle X_t^4 \rangle - 3 \langle X_t^2 \rangle^2 = 3g_4(\alpha,\beta) t^2$, where $g_4$ is a symmetric function given in SM and $g_4(1,\beta) = \beta^{-3}(\beta-1)^2 $. 
This function is shown FIG. \ref{fig:cumulants}(b) for $\alpha=\beta$.

\textit{Non-monotonicity and smoothness of $p_{\alpha,\beta}(u)$.} A second remarkable  feature of the propagator $P_{\alpha,\beta}(x,t)$,  which is not captured by a direct analysis of the moments of $X_t$, is its non-monotonicity. In fact, for some values of $\alpha, \beta < 1$ (repelling SATWs), we find that the distribution has local maxima, or bumps, for non-zero values of $x$, which we write as $x_{\pm}(t)$. While explicit exact expressions for $x_\pm$ seem out of reach, a good approximation can be obtained  for  $\alpha, \beta$ small by keeping only the first two terms in the sum \eqref{propag} defining $p_{\alpha,\beta}(u)$. Solving $\partial_u p_{\alpha,\beta}(u) = 0$ with this approximation  then yields
\begin{equation}
\label{approx-bump}
    x_+(t) \Sim{\alpha, \beta \to 0} \sqrt{\frac{t}{2(1+\beta)} \log \frac{(1-\alpha) (1-\beta) (\beta +2)^3}{\beta^2 (\beta +3) (\alpha +2 \beta +1)}} ;
\end{equation}
 a similar expression for $x_-$ is obtained by swapping $\alpha$ and $\beta$. Requiring that the argument of  the $\log$ in this expression is greater than $1$ provides an explicit  condition on $\alpha,\beta$ for $P_{\alpha,\beta}(x,t)$ to have bumps for $x>0$ ;  in the SM, we verify that this  condition accurately predicts non monotonicity of the propagator for a broad range of values  of $\alpha, \beta$. The accuracy of the approximate prediction  \eqref{approx-bump} is illustrated in FIG.\ref{fig:pu}. This shows that self-repulsion at the edges of the visited territory, if strong enough, can drive the RWer away from 0 and yield a non monotonic propagator, while preserving the diffusive scaling. \par 
We now turn to regularity properties of the scaling function $p_{\alpha,\beta}$. First, note that due to the different definitions for $u>0$ and $u<0$ in \eqref{propag} for $\alpha\not=\beta$, writing the continuity of $p_{\alpha,\beta}$  at $u=0$  yields a non trivial identity involving  hypergeometric functions, which we indeed verify up to arbitrary precision numerically (see SM). Second, we show in SM (see  FIG. \ref{fig:pu}) that $p_{\alpha,\beta}$  has a continuous,  vanishing derivative at $u=0$ for all $\alpha, \beta$ : this is in contrast with the propagator  of the true self-avoiding walk \cite{dumaz}, or locally activated random walks \cite{larw-pre,larw-new}, which have  singular derivatives at $u=0$. 

\textit{Extension: polynomially self-repelling walk (PSRW).} Last, we obtain the exact propagator of the PSRW introduced in \cite{toth96}. It is defined as a RW on $\mathbb{Z}$ with jump probabilities
\begin{equation}
    P_t(x + 1|x) = \frac{w(L_t(x+1))}{w(L_t(x+1) + w(L_t(x))} = 1-P_t(x-1|x)
\end{equation}
where $L_t(x)$ is the number of times the RWer has crossed the unoriented edge $\{x,x- 1\}$ up to time $t$, and $w(n) \Sim{n \to \infty} n^{-\gamma}$ for  $0<\gamma<\infty$. Although so far no explicit expression could be obtained for the propagator of the PSRW, it was shown in \cite{toth96}  that the scaling variable $u = x / \sqrt{2t(2\gamma+1)}$, in the regime $x,t \to \infty, u$ fixed, is distributed according to $p_{1/2,1/2}(u)$. Our explicit expression \eqref{propag} thus provides readily an exact expression of the propagator of the PSRW as a by-product.  Numerical simulations FIG. \ref{fig:simus}(b) confirm this  result. In particular,  an exact determination of the  diffusion coefficient of the PSRW is deduced from our result \eqref{D} for $\alpha=\beta=1/2$ and reads 
\begin{equation}
    D_{PSRW} = (2\gamma+1) \left(\frac{1}{2} + C + \frac{\pi^2}{16} \right),
\end{equation} 
where $C$ is Catalan's constant.

Finally, we have obtained an exact,  explicit expressions for the propagator of two important classes of strongly non Markovian random walks, namely, the once-reinforced random walk (SATW) and the polynomially self-repelling walk (PSRW). These analytic expressions provide benchmark  results which  give access to various observables of broad relevance, such as the diffusion coefficient, which has so far remained undetermined. Our results reveal remarkable features of the SATW, such as a smooth, non monotonic behaviour of the propagator in the case of strong enough self-repulsion induced by the   inherently non-Markovian nature of the dynamics.



\section*{Supplementary material}

# Supplementary Material - Exact densities of one-dimensional self-interacting random walks

April 24, 2024

## 1 Definition of hypergeometric functions

As hypergeometric functions will be used frequently in the following, we remind their definition. First, we define the Pochhammer symbol

$$(q)_n = q(q+1)\ldots(q+n-1) \tag{1}$$

If $a_1,\ldots,a_p,b_1,\ldots,b_q$ are suitable complex numbers, then

$$_pF_q(a_1,\ldots,a_p;b_1,\ldots,b_q;z) = \sum_{n=0}^{\infty} \frac{(a_1)_n\ldots(a_p)_n}{(b_1)_n\ldots(b_q)_n}\frac{z^n}{n!} \tag{2}$$

The regularized hypergeometric function is defined as

$$_p\tilde{F}_q(a_1,\ldots,a_p;b_1,\ldots,b_q;z) = \frac{_pF_q(a_1,\ldots,a_p;b_1,\ldots,b_q;z)}{\Gamma(b_1)\ldots\Gamma(b_q)} \tag{3}$$

## 2 Derivation of the quantities $F_L^{\alpha,\beta}$ and Eq. (5) of the main text

In this section we derive the expression of the matrices $F_L^{\alpha,\beta}$ as a function of $F_L^{1,1}$. We remind that $F_L^{\alpha,\beta}(\tau,\varepsilon|\varepsilon')$ is defined as the distribution of the exit time $\tau$ from the interval of size $L$ starting at boundary $\varepsilon' = \pm$ through the boundary $\varepsilon$. We start by writing a self-consistent equation for the quantities $F_L^{\alpha,\beta}(\tau,\varepsilon'|\varepsilon)$. The interval of visited sites is $[-m,k]$. Since jump probabilities are modified only at $-m,k$, in $[-m+1,k-1]$, the RWer behaves as a simple RWer. A simple partition over the first step of the RWer then allows us to write, if $\varepsilon' = +, \varepsilon = -$

$$F_L^{\alpha,\beta}(\tau,+|-) = \frac{\beta}{1+\beta}\sum_{t=0}^{\tau-1}\left[F_{L-2}^{1,1}(t,+|-)F_L^{\alpha,\beta}(\tau-1-t,+|+) + F_{L-2}^{1,1}(t,-|-)F_L^{\alpha,\beta}(\tau-1-t,+|-)\right] \tag{4}$$

and if $\varepsilon' = +, \varepsilon = +$

$$F_L^{\alpha,\beta}(\tau,+|+) = \frac{\delta_{\tau,1}}{1+\beta} + \frac{\beta}{1+\beta}\sum_{t=0}^{\tau-1}\left[F_{L-2}^{1,1}(t,+|+)F_L^{\alpha,\beta}(\tau-1-t,+|+) + F_{L-2}^{1,1}(t,-|+)F_L^{\alpha,\beta}(\tau-1-t,+|-)\right] \tag{5}$$

1

Other values of $\varepsilon, \varepsilon'$ can be deduced in the same way. Taking the generating functions on both sides and remarking the matrix product structure of the RHS yields

$$F_L^{\alpha,\beta}(\xi) = \begin{bmatrix} \xi/(1+\beta) & 0 \\ 0 & \xi/(1+\alpha) \end{bmatrix} + \xi F_L^{\alpha,\beta} F_{L-2}^{1,1} \begin{bmatrix} \beta/(1+\beta) & 0 \\ 0 & \alpha/(1+\alpha) \end{bmatrix} \tag{6}$$

The convention for matrix indices is $\begin{bmatrix} +|+ & +|- \\ -|+ & -|- \end{bmatrix}$. Finally, we find the expression for $F^{1,1}$ in [1]

$$\hat{F}_L^{1,1}(\xi) = \frac{1}{\sinh(L+2)\mu} \begin{bmatrix} \sinh(L+1)\mu & \sinh\mu \\ \sinh\mu & \sinh(L+1)\mu \end{bmatrix} \tag{7}$$

Using $\xi = \frac{1}{\cosh\mu}$, one finds the following exact expression from (6)

$$\hat{F}_L^{\alpha,\beta} = \frac{\text{sech}(\mu)}{\alpha\beta\tanh^2(\mu) + (\alpha+\beta)\tanh(\mu)\coth(\mu L) + 1} \left( \mathbb{1}_2 + \begin{bmatrix} \alpha\tanh\mu\coth(L\mu) & \alpha\frac{\tanh\mu}{\sinh(L\mu)} \\ \beta\frac{\tanh\mu}{\sinh(L\mu)} & \beta\tanh\mu\coth(L\mu) \end{bmatrix} \right) \tag{8}$$

from which we deduce the scaling form given in Eq. (5) of the main text, recalling that $\mu \underset{s\to 0}{\sim} \sqrt{2s}$.

## 3 Proof of the symmetry relation given by Eq. (3) of the main text

Here we prove that the symmetry relation

$$\frac{\hat{\pi}_{-m,\underline{k}}}{\hat{\pi}_{\underline{-m},k}} = \frac{\sinh(k\mu)}{\sinh(m\mu)} \tag{9}$$

given by Eq. (3) of the main text is exact. We remind that $\hat{\pi}_{-m,\underline{k}}(\xi) = \sum_{t=0}^{\infty} \xi^t \pi_{-m,\underline{k}}(t)$ is the generating function of the joint distribution $\pi_{-m,\underline{k}}(t)$ of the minimum $-m$ and of the time $t$ needed to reach the maximum $k$ We recall that this relation yields the following exact identities (using the space reversal identity $F^{\alpha,\beta}(\varepsilon'|\varepsilon) = F^{\beta,\alpha}(-\varepsilon'|-\varepsilon)$)

$$\hat{\pi}_{-m,\underline{k}} = F_m^{\alpha,\beta}(+|-) \prod_{j=0}^{m-1} F_j^{\alpha,\beta}(-|-) \times \prod_{j=1}^{k-1} \left[ \frac{\sinh m\mu}{\sinh j\mu} F_{j+m}^{\alpha,\beta}(+|-) + F_{j+m}^{\alpha,\beta}(+|+) \right]$$

$$\hat{\pi}_{\underline{-m},k} = F_k^{\beta,\alpha}(+|-) \prod_{j=0}^{k-1} F_j^{\beta,\alpha}(-|-) \times \prod_{j=1}^{m-1} \left[ \frac{\sinh k\mu}{\sinh j\mu} F_{k+j}^{\beta,\alpha}(+|-) + F_{k+j}^{\beta,\alpha}(+|+) \right] \tag{10}$$

where $\mu = -\log\left((1-\sqrt{1-\xi^2})/\xi\right)$. Using (8), one can write the following ratio, singling out the term $F_0^{\alpha,\beta}(+|+) = \xi\frac{\alpha}{\alpha+\beta}$

$$\frac{\hat{\pi}_{-m,\underline{k}}}{\hat{\pi}_{\underline{-m},k}} = \frac{\beta}{\alpha} \frac{\alpha\sinh(k\mu)}{\beta\sinh(m\mu)} \frac{\prod_{j=1}^{m-1}(1+\beta\tanh\mu\coth(j\mu))}{\prod_{j=1}^{k-1}(1+\alpha\tanh\mu\coth(j\mu))} \frac{\prod_{j=1}^{k-1}\left(\frac{\sinh m\mu}{\sinh j\mu}\frac{\alpha\tanh\mu}{\sinh(m+j)\mu} + 1 + \alpha\tanh\mu\coth(m+j)\mu\right)}{\prod_{j=1}^{m-1}\left(\frac{\sinh k\mu}{\sinh j\mu}\frac{\beta\tanh\mu}{\sinh(k+j)\mu} + 1 + \beta\tanh\mu\coth(k+j)\mu\right)} \tag{11}$$

2

Thus it remains to show

$$\frac{\prod_{j=1}^{m-1}(1+\beta\tanh\mu\coth(j\mu))\prod_{j=1}^{k-1}\left(\frac{\sinh m\mu}{\sinh j\mu}\frac{\alpha\tanh\mu}{\sinh(m+j)\mu}+1+\alpha\tanh\mu\coth(m+j)\mu\right)}{\prod_{j=1}^{k-1}(1+\alpha\tanh\mu\coth(j\mu))\prod_{j=1}^{m-1}\left(\frac{\sinh k\mu}{\sinh j\mu}\frac{\beta\tanh\mu}{\sinh(k+j)\mu}+1+\beta\tanh\mu\coth(k+j)\mu\right)} = 1 \quad (12)$$

The above holds because of the following relation

$$\coth(\mu(j+m)) + \frac{\sinh(\mu m)}{\sinh(j\mu)\sinh(\mu(j+m))} = \coth(j\mu) \quad (13)$$

This proves that (9) (given as Eq. (3) of the main text) is exact.

## 4 Scaling expression of $\pi_{\underline{k},-m}$ (Eq. (6) of main text)

Here we derive the diffusive $(k, m \gg 1, s \to 0, (k+m)\sqrt{s}$ fixed) expression of $\pi_{\underline{k},-m}$ given by Eq. (6) of the main text. We start from the exact identity

$$\pi_{-m,\underline{k}} = F_m^{\alpha,\beta}(+|-) \prod_{j=0}^{m-1} F_j^{\alpha,\beta}(-|-) \times \prod_{j=1}^{k-1}\left[\frac{\sinh m\sqrt{2s}}{\sinh j\sqrt{2s}} F_{j+m}^{\alpha,\beta}(+|-) + F_{j+m}^{\alpha,\beta}(+|+)\right] \quad (14)$$

and use the asymptotics

$$F_L^{\alpha,\beta} \sim \mathbb{1}_2 + \sqrt{2s}\begin{bmatrix} -\beta\coth(L\sqrt{2s}) & \frac{\alpha}{\sinh L\sqrt{2s}} \\ \frac{\beta}{\sinh L\sqrt{2s}} & -\alpha\coth(L\sqrt{2s}) \end{bmatrix} \quad (15)$$

This allows us to rewrite in the scaling regime, provided the integrals below all diverge as $k, m \to \infty$ (this is checked a posteriori)

$$\pi_{-m,\underline{k}} \sim \frac{C(\alpha,\beta)\sqrt{2s}}{\sinh m\sqrt{2s}} \exp\left(\int_1^m \log\left(1 - \alpha\sqrt{2s}\coth\left(j\sqrt{2s}\right)\right)dj\right) \times$$
$$\exp\left(\int_1^k \log\left[1 + \frac{\sinh m\sqrt{2s}}{\sinh(k-j)\sqrt{2s}}\frac{\alpha\sqrt{2s}}{\sinh(k-j+m)\sqrt{2s}} - \sqrt{2s}\beta\coth\left((k-j+m)\sqrt{2s}\right)\right]dj\right) \quad (16)$$

where $C(\alpha,\beta)$ is a constant, that stems from the approximation of the integrands by their asymptotics. In the scaling regime, we can simplify the above to

$$\pi_{-m,\underline{k}} \sim \frac{C(\alpha,\beta)\sqrt{2s}}{\sinh m\sqrt{2s}} \exp\left(-\int_1^m \alpha\sqrt{2s}\coth\left(j\sqrt{2s}\right)dj\right) \times$$
$$\exp\left(\int_1^k \left[\frac{\sinh m\sqrt{2s}}{\sinh(k-j)\sqrt{2s}}\frac{\alpha\sqrt{2s}}{\sinh(k-j+m)\sqrt{2s}} - \sqrt{2s}\beta\coth\left((k-j+m)\sqrt{2s}\right)\right]dj\right) \quad (17)$$

3

One can compute the following integrals

$$\sqrt{2s}\int_1^m \coth\!\left(j\sqrt{2s}\right)dj = \log\!\left(\frac{\sinh m\sqrt{2s}}{\sinh\sqrt{2s}}\right)$$

$$\sqrt{2s}\int_1^k \coth\!\left((k-j+m)\sqrt{2s}\right)dj = \log\!\left(\frac{\sinh(k+m)\sqrt{2s}}{\sinh m\sqrt{2s}}\right)$$

$$\sqrt{2s}\sinh\!\left(m\sqrt{2s}\right)\int_1^k \frac{dj}{\sinh(k-j)\sqrt{2s}\sinh(k-j+m)\sqrt{2s}} = -\log\!\left(\frac{\sinh\!\left(\sqrt{2s}\right)\sinh\!\left(\sqrt{2s}(k+m)\right)}{\sinh\!\left(k\sqrt{2s}\right)\sinh\!\left((m+1)\sqrt{2s}\right)}\right) \quad (18)$$

Hence, we obtain

$$\pi_{-m,\underline{k}} \sim C(\alpha,\beta)\frac{\sqrt{2s}}{\sinh\!\left(m\sqrt{2s}\right)}\frac{\sinh^\beta(m\sqrt{2s})\sinh^\alpha(k\sqrt{2s})}{\sinh^{\alpha+\beta}((k+m)\sqrt{2s})} \quad (19)$$

Now, to obtain the constant $C(\alpha,\beta)$, we obtain a simple and exact formula for $\pi_{-m,\underline{k}}(s=0)$. We will then match the behavior of the latter quantity with (19) in the scaling regime. To obtain $\pi_{-m,\underline{k}}(s=0)$ we first compute the first-passage matrices at $s=0$. We find from (8) by taking the limit $\mu \underset{s\to 0}{\sim} \sqrt{2s} \to 0$

$$F_L^{\alpha,\beta}(s=0) = \frac{1}{L+\alpha+\beta}\begin{bmatrix} L+\alpha & \alpha \\ \beta & L+\beta \end{bmatrix} \quad (20)$$

Hence, at $s=0$, (14) becomes

$$\pi_{-m,\underline{k}}(s=0) = \frac{\alpha}{m+\alpha+\beta}\prod_{j=0}^{m-1}\frac{j+\beta}{j+\alpha+\beta} \times \prod_{j=1}^{k-1}\left[\frac{m}{k-j}\frac{\alpha}{k-j+m+\alpha+\beta} + \frac{k-j+m+\alpha}{k-j+m+\alpha+\beta}\right] \quad (21)$$

This simplifies to

$$\pi_{-m,\underline{k}}(s=0) = \frac{\alpha}{m+\alpha+\beta}\prod_{j=0}^{m-1}\frac{j+\beta}{j+\alpha+\beta} \times \prod_{j=1}^{k-1}\frac{(k-j+m)(k-j+\alpha)}{(k-j)(k-j+m+\alpha+\beta)} \quad (22)$$

Writing it more clearly

$$\pi_{-m,\underline{k}}(s=0) = \frac{(m+k-1)\ldots(m+1)}{(k-1)\ldots 1} \times \frac{\beta\ldots(\beta+m-1)\times\alpha\ldots(\alpha+k-1)}{(\beta+\alpha)\ldots(\beta+\alpha+k+m-1)} \quad (23)$$

We recognize the Beta function as well as a binomial coefficient. We thus obtain the exact formulae, using $\frac{\pi_{-m,\underline{k}}}{\pi_{\underline{-m},k}} = \frac{k}{m}$

$$\pi_{-m,\underline{k}}(s=0) = \binom{k+m-1}{k-1}\frac{B(k+\alpha,m+\beta)}{B(\alpha,\beta)},\ \pi_{\underline{-m},k}(s=0) = \binom{k+m-1}{k}\frac{B(k+\alpha,m+\beta)}{B(\alpha,\beta)}, \quad (24)$$

Matching (24) with (19) (which is possible since (24) is exact), one obtains finally that $C(\alpha,\beta) = \frac{1}{B(\alpha,\beta)}$. Eq. (19) above then yields Eq.(6) of the main text.

4

# 5 Exact splitting probability of the SATW

As an illustration of the usefulness of the $\pi_{-j,\underline{k}}$, which show that they yield classical splitting probabilities, which so far remained unknown for the SATW. Define the probability $Q_+(k,m,t)$ that the RWer reaches a target site $k$ for the first time at time $t$ before reaching another target site $-m$. Using the expression

$$\mathbb{P}(\text{reach } k \text{ before } -m) = \sum_{t=0}^{\infty} Q_+(k,m,t) = \sum_{j=0}^{m-1} \hat{\pi}_{-j,\underline{k}}(s=0) \tag{25}$$

we deduce the exact splitting probability $\mathbb{P}(\text{reach } k \text{ before } -m)$ of the SATW, which is the probability to have reached site $k$ before site $-m$ starting from 0 at $t=0$.

$$\mathbb{P}(\text{reach } k \text{ before } -m) = 1 - \frac{\Gamma(k+\alpha)\Gamma(k+m)\Gamma(m+\beta)\,_3F_2(1,k+m,m+\beta;m+1,k+m+\alpha+\beta;1)}{B(\alpha,\beta)\Gamma(k)\Gamma(m+1)\Gamma(k+m+\alpha+\beta)} \tag{26}$$

# 6 Propagator in the span with absorption at the span boundaries (Eq. (8) of the main text)

In this section we compute exactly the probability $P_s^{m,k}(x,\tau|\varepsilon)$ that the RWer is at site $x$ at time $t+\tau$ and has remained within the interval $[-m,k]$, which is the span of the RW at time $t$. To be on site $x$ at time $t+\tau$ starting from, say, $-m$, there are three possible scenarii. First, if $x=-m$, $\tau$ is allowed to be 0. If $x\neq -m$, it means that the RWer jumped from $-m$ to reach $x$. Did it go from $-m$ to $x$ at time $t+\tau$ without reaching $-m$ again or reaching $k$ ? If so, it behaves exactly like a simple RWer in the interval $[-m+1, k-1]$. Now, assume it did reach one of these sites : say it reached $-m$ first. Then, the process begins anew, starting from $-m$. This renewal approach yields the following identities in generating functions (introducing $\hat{P}_{s,\varepsilon}(x) = \hat{P}_s^{m,k}(x,\xi|\varepsilon)$)

$$\hat{P}_{s,-}(x) = \delta_{x,-m} + \xi\frac{\alpha}{1+\alpha}\left(\hat{P}_{\text{polya},s}^{m-1,k-1}(x|-m+1) + \hat{F}_{\text{polya},s}^{L-2}(-|-)\hat{P}_{s,-}^{m-1,k-1}(x) + F_{\text{polya},s}^{L-2}(+|-)\hat{P}_{s,+}^{m-1,k-1}(x)\right)$$

$$\hat{P}_{s,+}(x) = \delta_{x,k} + \xi\frac{\beta}{1+\beta}\left(\hat{P}_{\text{polya},s}^{m-1,k-1}(x|k-1) + F_{\text{polya},s}^{L-2}(+|+,\xi)P_{s,+}^{m,k}(x) + F_{\text{polya},s}^{L-2}(-|+)P_{s,-}^{m,k}(x)\right) \tag{27}$$

where the quantities with index 'polya' refer to the quantities for the simple random walk $\alpha=\beta=1$. We can write this in a matrix form, introducing the ket $\left|\hat{P}_s(x)\right\rangle = \left|\hat{P}_{s,-}(x), \hat{P}_{s,+}(x)\right\rangle$

$$\left|\hat{P}_s(x)\right\rangle = \left|\delta_{x,-a} + \frac{\xi\alpha(1-\delta_{x,-a})}{1+\alpha}\hat{P}_{\text{polya},s}^{L-2}(x|-a+1,\xi), \delta_{x,b} + \frac{\xi\beta(1-\delta_{x,b})}{1+\beta}\hat{P}_{\text{polya},s}^{L-2}(x|b-1,\xi)\right\rangle + \xi\begin{bmatrix}\frac{\alpha}{1+\alpha}\hat{F}_{\text{polya}}^{L-2}(-|-) & \frac{\alpha}{1+\alpha}\hat{F}_{\text{polya}}^{L-2}(+|-) \\ \frac{\beta}{1+\beta}\hat{F}_{\text{polya}}^{L-2}(-|+) & \frac{\beta}{1+\beta}\hat{F}_{\text{polya}}^{L-2}(+|+)\end{bmatrix}\left|\hat{P}_s(x)\right\rangle \tag{28}$$

This yields

$$\hat{P}_{s,-}(x \neq -m, k) =$$
$$\frac{2\alpha\operatorname{csch}(\mu L)(\beta \tanh(\mu)\operatorname{csch}(\mu L) \sinh(\mu(a+x)) + \sinh(\mu(b-x))(\beta \tanh(\mu) \coth(\mu L) + 1))}{\alpha\beta \tanh^2(\mu) + (\alpha+\beta) \tanh(\mu) \coth(\mu L) + 1}$$
$$\hat{P}_{s,+}(x \neq -m, k) =$$
$$\frac{2\beta\operatorname{csch}(\mu L)(\sinh(\mu(a+x))(\alpha \tanh(\mu) \coth(\mu L) + 1) + \alpha \tanh(\mu)\operatorname{csch}(\mu L) \sinh(\mu(b-x)))}{\alpha\beta \tanh^2(\mu) + (\alpha+\beta) \tanh(\mu) \coth(\mu L) + 1}$$
(29)

In the limit $k, m \gg 1, s \to 0, L\sqrt{s}$ fixed, these simplify to

$$\hat{P}_{s,-}(x) = 2\alpha \frac{\sinh(k-x)\mu}{\sinh L\mu}, \hat{P}_{s,+}(x) = 2\beta \frac{\sinh(m+x)\mu}{\sinh L\mu} \quad (30)$$

which give Eq. (8) of the main text.

## 7 Computing the scaling function $p_{\alpha,\beta}$ (Eq. (1) of the main text)

We remind that we have the following expression for the joint distribution of min, max and position in the scaling limit, obtained as Eq. (9) of the main text

$$\hat{Q}(k, m, x, s) \sim \frac{2\sqrt{2s}}{B(\alpha, \beta)} \frac{\sinh^\beta(m\sqrt{2s}) \sinh^\alpha(k\sqrt{2s})}{\sinh^{\alpha+\beta+1}((k+m)\sqrt{2s})} \left( \beta \frac{\sinh(m+x)\sqrt{2s}}{\sinh(m\sqrt{2s})} + \alpha \frac{\sinh(k-x)\sqrt{2s}}{\sinh(k\sqrt{2s})} \right) \quad (31)$$

Taking the marginal of the above gives the Laplace transform of the propagator in the scaling limit (in the half space $x > 0$)

$$\hat{P}_{\alpha,\beta}(x, s) = \int_0^\infty P_{\alpha,\beta}(x, t)e^{-st} dt \sim$$
$$\frac{2\sqrt{2s}}{B(\alpha, \beta)} \int_0^\infty dm \int_x^\infty dk \frac{\sinh^\beta m\sqrt{2s} \sinh^\alpha k\sqrt{2s}}{\sinh^{\alpha+\beta+1}(k+m)\sqrt{2s}} \left( \alpha \frac{\sinh(k\sqrt{2s} - x\sqrt{2s})}{\sinh k\sqrt{2s}} + \beta \frac{\sinh(m\sqrt{2s} + x\sqrt{2s})}{\sinh m\sqrt{2s}} \right) \quad (32)$$

Writing $u = x\sqrt{2s}$ and rescaling the integration variables $a = m\sqrt{2s}, b = k\sqrt{2s}$, it simply remains to compute the following quantity

$$\Phi^{\alpha,\beta}(u) = \frac{1}{B(\alpha, \beta)} \int_0^\infty da \int_u^\infty db \frac{\sinh^\beta a \sinh^\alpha b}{\sinh^{\alpha+\beta+1}(a+b)} \left( \alpha \frac{\sinh(b-u)}{\sinh b} + \beta \frac{\sinh(a+u)}{\sinh a} \right) \quad (33)$$

We now present our strategy to compute exactly this double integral.

### 7.1 Important identities

Several important identites can be readily obtained by Mathematica.

$$\int_0^\infty \frac{\sinh^c a}{\sinh^d(a+b)} da =$$
$$e^{-bd} 2^{-c+d-1} B\left(c+1, \frac{d-c}{2}\right) {}_2F_1\left(d, \frac{d-c}{2}; \frac{1}{2}(c+d+2); e^{-2b}\right) \quad (34)$$

6

$$\int_0^\infty \frac{\sinh^c a}{\sinh^d(a+b)} \coth(a) da =$$
$$e^{-bd} 2^{-c+d-1} \left[ B\left(c, \frac{d-c}{2}\right) {}_2F_1\left(d, \frac{d-c}{2}; \frac{c+d}{2}; e^{-2b}\right) + B\left(c, \frac{2+d-c}{2}\right) {}_2F_1\left(d, \frac{d-c+2}{2}; \frac{c+d+2}{2}; e^{-2b}\right) \right] \tag{35}$$

We recall the Euler transformation of the hypergeometric function ${}_2F_1$

$${}_2F_1(a,b,c;z) = (1-z)^{c-a-b} {}_2F_1(c-a, c-b, c; z) \tag{36}$$

Using this transformation (and relations on the Beta function $B$) allows us to rewrite the integrals above as

$$\int_0^\infty \frac{\sinh^c a}{\sinh^d(a+b)} da = \frac{e^{-b(1+c)}}{\sinh^{d-c-1}(b)} \frac{2c}{c+d} B\left(c, \frac{d-c}{2}\right) {}_2F_1\left(1+\frac{c-d}{2}, 1+c; 1+\frac{c+d}{2}; e^{-2b}\right) \tag{37}$$

$$\int_0^\infty \frac{\sinh^c a}{\sinh^d(a+b)} \coth(a) da =$$
$$\frac{e^{-bc}}{2\sinh^{d-c}(b)} B\left(c, \frac{d-c}{2}\right) \left[ {}_2F_1\left(\frac{c-d}{2}, c; \frac{c+d}{2}; e^{-2b}\right) + \frac{d-c}{c+d} {}_2F_1\left(1+\frac{c-d}{2}, c; 1+\frac{c+d}{2}; e^{-2b}\right) \right] \tag{38}$$

The Gauss contiguous relations for the hypergeometric ${}_2F_1$ function imply

$$aF(a+) = \frac{(1+a-c)F + (c-1)F(b-,c-)}{1-z} \iff F(b-,c-) = \frac{(1-z)aF(a+) + (c-a-1)F}{c-1} \tag{39}$$

where $F = {}_2F_1(a,b,c;z), F(a+) = {}_2F_1(a+1,b,c;z), \ldots$

In particular, if $F = {}_2F_1(1+\frac{c-d}{2}, c, 1+\frac{c+d}{2}; e^{-2b})$, $F_+ = {}_2F_1(1+\frac{c-d}{2}, 1+c, 1+\frac{c+d}{2}; e^{-2b})$, ie $a \mapsto c, b \mapsto 1+\frac{c-d}{2}, c \mapsto 1+\frac{c+d}{2}$, we obtain the following

$$\int_0^\infty \frac{\sinh^c a}{\sinh^d(a+b)} \coth(a) da =$$
$$\frac{e^{-bc}}{2(c+d)\sinh^{d-c}(b)} B\left(c, \frac{d-c}{2}\right) \left[ 4ce^{-b} \sinh(b) F_+ + 2(d-c) F \right] \tag{40}$$

Finally, there is a surprising summation formula for ratios of Pochhammer symbols. First, notice that if we define

$$c_k = \frac{\left(\frac{1-\alpha}{2}\right)_k (\beta)_k}{k! \left(\frac{1}{2}(\alpha+2\beta+3)\right)_k} \tag{41}$$

then one has the telescoping relation

$$(\alpha+k)(2k+1-\beta)c_k - (\alpha+k-1)(2k-1-\beta)c_{k-1} = (\beta - \alpha(\beta+2k))c_k \tag{42}$$

Thus, the following sum can be evaluated exactly (note that other similar sums cannot be evaluated exactly in terms of simple functions)

$$\sum_{k=0}^n (\beta - \alpha(\beta+2k))c_k = (-\alpha+2n+1)(\beta+n)c_n \tag{43}$$

7

## 7.2 Computing $\Phi^{\alpha,\beta}$

Write
$$\eta_1 = \int_0^\infty \frac{\sinh^\beta a}{\sinh^{\alpha+\beta+1}(a+b)} da, \eta_2 = \int_0^\infty \frac{\sinh^\beta a}{\sinh^{\alpha+\beta+1}(a+b)} \frac{\sinh(a+u)}{\sinh a} da \tag{44}$$

Notice that $\frac{\sinh(a+u)}{\sinh a} = \cosh(u) + \sinh(u)\coth(a)$. This way we can write

$$\eta_2 = \cosh(u)\eta_1 + \sinh(u)\psi, \psi = \int_0^\infty \frac{\sinh^\beta a}{\sinh^{\alpha+\beta+1}(a+b)} \coth(a) da \tag{45}$$

so that, for **positive $x$ or $u > 0$**

$$\Phi^{\alpha,\beta}(u) = \frac{1}{B(\alpha,\beta)} \int_0^\infty da \int_u^\infty db \frac{\sinh^\beta a \sinh^\alpha b}{\sinh^{\alpha+\beta+1}(a+b)} \left( \alpha \frac{\sinh(b-u)}{\sinh b} + \beta \frac{\sinh(a+u)}{\sinh a} \right) = \frac{1}{B(\alpha,\beta)} \int_u^\infty \eta(b) db$$

$$\eta = \sinh^\alpha(b) \left[ \eta_1 \left( \beta \cosh(u) + \alpha \frac{\sinh(b-u)}{\sinh(b)} \right) + \beta \sinh(u)\psi \right]$$

$$\eta = \sinh^\alpha(b) \left[ \eta_1 \left[ (\alpha+\beta) \cosh(u) - \alpha \sinh(u)\coth(b) \right] + \beta \sinh(u)\psi \right] \tag{46}$$

Define $F = {}_2F_1(\frac{1-\alpha}{2}, \beta, \frac{3+2\beta+\alpha}{2}; e^{-2b}), F_+ = {}_2F_1(\frac{1-\alpha}{2}, \beta+1, \frac{3+2\beta+\alpha}{2}; e^{-2b})$. Then one has, according to the identities we already showed above

$$\eta_1 \sinh^\alpha(b) = \frac{2\beta}{2\beta+\alpha+1} e^{-b(\beta+1)} B\left( \beta, \frac{1+\alpha}{2} \right) F_+$$

$$\psi \sinh^\alpha(b) = \frac{e^{-b\beta}}{1+2\beta+\alpha} B\left( \beta, \frac{1+\alpha}{2} \right) \left[ 2\beta e^{-b} F_+ + (1+\beta) \frac{F}{\sinh b} \right] \tag{47}$$

One thus finds

$$\boxed{\eta = \frac{\beta e^{-b(\beta+1)} B\left(\beta, \frac{\alpha+1}{2}\right)}{\alpha+2\beta+1} \left( \sinh(u)\big(\coth(b)((\alpha+1)F - 2\alpha F_+) + (\alpha+1)F + 2\beta F_+\big) + 2(\alpha+\beta)\cosh(u)F_+ \right)} \tag{48}$$

Now, let

$$c_n = \frac{\left(\frac{1-\alpha}{2}\right)_n (\beta)_n}{n! \left(\frac{3+2\beta+\alpha}{2}\right)_n}, \text{ so that } \frac{\left(\frac{1-\alpha}{2}\right)_n (\beta+1)_n}{n! \left(\frac{3+2\beta+\alpha}{2}\right)_n} = \frac{\beta+n}{\beta} c_n \tag{49}$$

these are the coefficients of the hypergeometric functions we are using

$$F = \sum_{n=0}^\infty c_n e^{-2nb}, F_+ = \frac{1}{\beta} \sum_{n=0}^\infty (\beta+n) c_n e^{-2nb} \tag{50}$$

We will have to perform the Cauchy product of these hypergeometric functions with the series representation of $\frac{1}{\sinh b} = 2e^{-b} \sum_{k=0}^\infty e^{-2kb}$, or equivalently $\coth b = 1 + \frac{e^{-b}}{\sinh b} = 1 + 2e^{-2b} \sum_{k=0}^\infty e^{-2kb}$. The term that is difficult to integrate with respect to $b$ is $(\coth(b) - 1)((\alpha+1)F - 2\alpha F_+)$. It writes, after a Cauchy product

$$(\coth(b) - 1)((\alpha+1)F - 2\alpha F_+) = \frac{2e^{-2b}}{\beta} \sum_{n=0}^\infty e^{-2\beta n} \sum_{k=0}^n (\beta(\alpha+1) - 2\alpha(\beta+k)) c_k \tag{51}$$

8

We identify in the RHS the sum (43) calculated above, which yields

$$(\coth(b) - 1)((\alpha+1)F - 2\alpha F_+) = \frac{2e^{-2b}}{\beta} \sum_{n=0}^{\infty} e^{-2bn}(1 + 2n - \alpha)(n + \beta)c_n \tag{52}$$

Thus we have

$$\eta = \frac{2B\left(\beta, \frac{\alpha+1}{2}\right)}{\alpha + 2\beta + 1} \times$$

$$\sum_{n=0}^{\infty} e^{-b(2n+\beta+1)} c_n \left(\sinh(u)\left(\beta^2 + \beta + e^{-2b}(-\alpha + 2n + 1)(\beta + n) - \alpha n + \beta n\right) + (\alpha + \beta)(\beta + n)\cosh(u)\right) \tag{53}$$

This is easily integrated with respect to $b$, and yields

$$\Phi^{\alpha,\beta} = \frac{B\left(\beta, \frac{\alpha+1}{2}\right)}{(\alpha + 2\beta + 1)B(\alpha, \beta)} \times$$

$$\sum_{n=0}^{\infty} c_n e^{-u(\beta+2n+4)} \left(\frac{(e^{2u} - 1)(-\alpha + 2n + 1)(\beta + n)}{\beta + 2n + 3} + \frac{e^{2u}\left(2n\left(\alpha + \beta e^{2u}\right) + \beta\left(\alpha + e^{2u}(\alpha + 2\beta + 1) - 1\right)\right)}{\beta + 2n + 1}\right) \tag{54}$$

Finally, regrouping equal exponential terms, we can rewrite this as

$$\Phi^{\alpha,\beta} = \frac{(\beta - 1)B\left(\beta, \frac{\alpha+1}{2}\right)}{2B(\alpha, \beta)} \sum_{n=0}^{\infty} e^{-u(\beta+2n)} \frac{\left(\frac{1-\alpha}{2}\right)_n (\beta)_n}{n!\left(\frac{1+2\beta+\alpha}{2}\right)_n} \frac{n + \frac{\beta}{2}}{(n + \frac{\beta-1}{2})(n + \frac{\beta+1}{2})} \tag{55}$$

Now, we deduce the propagator in the scaling limit by Laplace inverting the above term-by-term. Thus we obtain exactly the scaling function $p_{\alpha,\beta}$ defined by $P_{\alpha,\beta}(x,t) \sim p_{\alpha,\beta}\left(u = \frac{x}{\sqrt{2t}}\right)/\sqrt{2t}$ in the region $x > 0$

$$\boxed{p_{\alpha,\beta}\left(u = \frac{x}{\sqrt{2t}}\right) = \frac{(\beta - 1)B\left(\beta, \frac{\alpha+1}{2}\right)}{B(\alpha, \beta)\sqrt{\pi}} \sum_{n=0}^{\infty} \frac{\left(\frac{1-\alpha}{2}\right)_n (\beta)_n}{\left(\frac{1+2\beta+\alpha}{2}\right)_n} \frac{n + \frac{\beta}{2}}{(n + \frac{\beta-1}{2})(n + \frac{\beta+1}{2})} \frac{e^{-u^2(2n+\beta)^2}}{n!}} \tag{56}$$

Remember that $P_{\alpha,\beta}(-x,t) = P_{\beta,\alpha}(x,t)$. Using this identity, we obtain the scaling function for all real values of $u$. This is Eq. (1) of the main text. Note that we can explicitly compute $\int_\mathbb{R} p_{\alpha,\beta}(u)du = 1$ : the scaling function is normalized, as it should.

## 8 Moments of the position $X_t$

In the following we will use the following decomposition $\langle X_t^n \rangle = \langle X_t^n \rangle_+ + \langle X_t^n \rangle_-$

$$\langle X_t^n \rangle_+ = \int_0^\infty x^n P_{\alpha,\beta}(x,t)dx, \quad \langle X_t^n \rangle_- = \int_{-\infty}^0 x^n P_{\alpha,\beta}(x,t)dx = (-1)^n \int_0^\infty x^n P_{\beta,\alpha}(x,t)dx \tag{57}$$

### 8.1 Average $\langle X_t \rangle$

Using

$$\int_0^\infty x \frac{e^{-\frac{x^2(2n+\beta)^2}{2t}}}{\sqrt{2\pi t}} dx = \frac{\sqrt{t}}{\sqrt{2\pi}(2n + \beta)^2} \tag{58}$$

9

we can explicitly compute the mean. In the scaling limit, one has, using our exact result for the propagator and the identity above

$$\frac{\langle X_t \rangle_+}{\sqrt{2t}} \sim k_1(\alpha, \beta) = \frac{2^{-\alpha}(\beta-1)\Gamma\left(\frac{\beta+1}{2}\right)\Gamma(\alpha+\beta)\,_3\tilde{F}_2\left(\frac{1-\alpha}{2}, \frac{\beta+1}{2}, \beta; \frac{\beta+3}{2}, \frac{\alpha+1}{2}+\beta; 1\right)}{\Gamma\left(\frac{\alpha}{2}\right)} + \frac{\alpha}{\sqrt{\pi}(\alpha+\beta)}$$
$$- \frac{\Gamma\left(\frac{\alpha+1}{2}\right)\Gamma\left(\frac{\beta}{2}\right)\Gamma\left(\frac{\alpha+\beta}{2}\right)}{\Gamma\left(\frac{\alpha}{2}\right)\Gamma\left(\frac{\beta-1}{2}\right)\Gamma\left(\frac{1}{2}(\alpha+\beta+1)\right)}$$
(59)

From the equality $\langle X_t \rangle = \langle X_t \rangle_+ - \langle X_t \rangle_- = \sqrt{2t}g_1(\alpha,\beta)$, we deduce

$$g_1(\alpha, \beta) = k_1(\alpha, \beta) - k_1(\beta, \alpha) \tag{60}$$

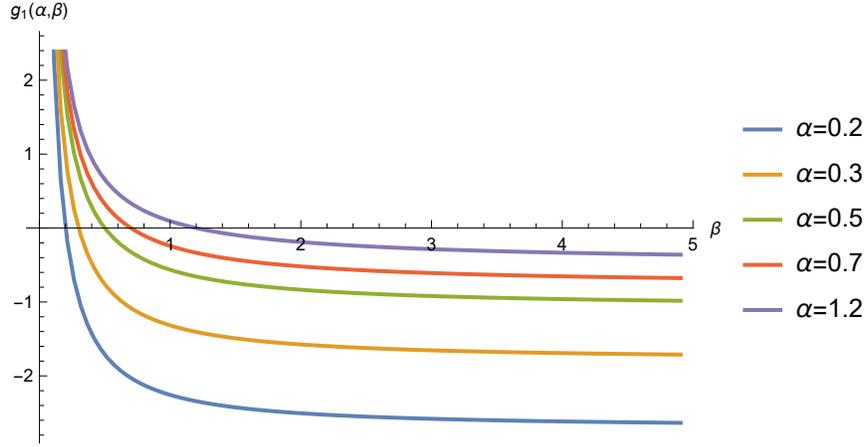

Figure 1: The function $g_1(\alpha, \beta) \sim \frac{\langle X_t \rangle}{\sqrt{2t}}$ plotted for different values of $\alpha$. As was expected, it becomes negative for $\beta > \alpha$, ie when the RWer is more likely to increment its span to the left than to the right.

## 8.2 Diffusion coefficient $D_{\alpha,\beta}$

### 8.2.1 Second moment $\langle X_t^2 \rangle$

Using

$$\int_0^\infty x^2 \frac{e^{-\frac{x^2(2n+\beta)^2}{2t}}}{\sqrt{2\pi t}} dx = \frac{t}{16(\beta/2+n)^3} \tag{61}$$

10

we explicitly compute the second moment. Following the same notation as before, we introduce $\langle X_t^2 \rangle_+ = \int_0^\infty x^2 P_{\alpha,\beta}(x,t) dx$ and we find

$$\frac{\langle X_t^2 \rangle_+}{t} \sim k_2(\alpha, \beta) = \frac{\alpha}{\alpha + \beta} - \frac{\sqrt{\pi} 2^{-\alpha-1} (\beta-1) \Gamma\left(\frac{\beta}{2}\right)^2 \Gamma(\alpha+\beta) \, {}_4\tilde{F}_3\left(\frac{1-\alpha}{2}, \frac{\beta}{2}, \frac{\beta}{2}, \beta; \frac{\beta+2}{2}, \frac{\beta+2}{2}, \frac{\alpha+1}{2} + \beta; 1\right)}{\Gamma\left(\frac{\alpha}{2}\right)} \tag{62}$$

Thus, the second moment writes

$$\frac{\langle X_t^2 \rangle}{t} = \frac{1}{t}\left(\langle X_t^2 \rangle_+ + \langle X_t^2 \rangle_-\right) = k_2(\alpha, \beta) + k_2(\beta, \alpha) \tag{63}$$

### 8.2.2 Variance

The variance can now be readily written

$$\frac{V(X_t)}{t} = \frac{1}{t}\left(\langle X_t^2 \rangle - \langle X_t \rangle^2\right) = k_2(\alpha, \beta) + k_2(\beta, \alpha) - 2(k_1(\alpha, \beta) - k_1(\beta, \alpha))^2 = 2 D_{\alpha,\beta} \tag{64}$$

No simplification seems to arise in the expression of $D_{\alpha,\beta}$ for general $\alpha, \beta$. However, in the symmetric case $\alpha = \beta$, we find the simpler expression given in the main text.

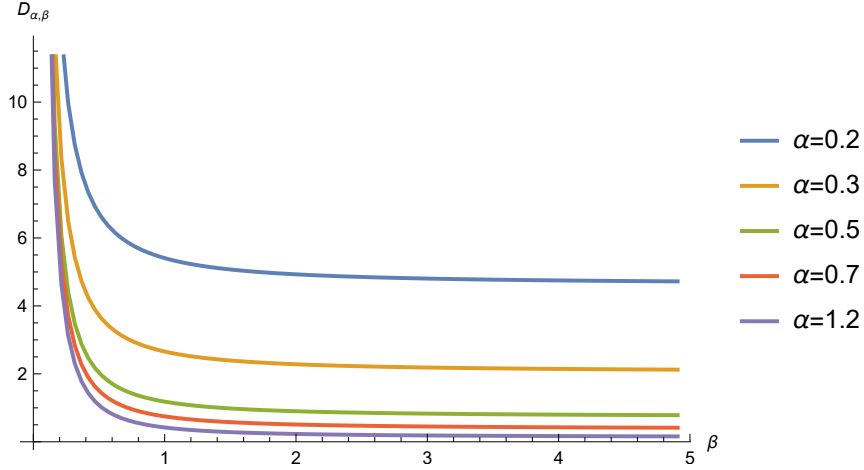

Figure 2: The diffusion coefficient $D_{\alpha,\beta}$ plotted for different values of $\alpha$.

## 8.3 Fourth cumulant

### 8.3.1 Fourth moment

The following integral is easily computed

$$\int_0^\infty x^4 \frac{e^{-\frac{x^2(\beta+2n)^2}{2t}}}{\sqrt{2\pi t}} dx = \frac{3t^2}{2^6(n+\beta/2)^5} \tag{65}$$

11

We find

$$\frac{\langle X_t^4\rangle_+}{3t^2} \sim k_4(\alpha,\beta) = \frac{\alpha}{\alpha+\beta} - \frac{\alpha\sqrt{\pi}2^{-\alpha}(\beta-1)\Gamma(\alpha+\beta)}{\beta^4\Gamma\left(\frac{\alpha}{2}+1\right)\Gamma\left(\frac{\alpha+1}{2}+\beta\right)} \times \Big[$$
$$\beta^2\,{}_4F_3\left(\frac{1}{2}-\frac{\alpha}{2},\frac{\beta}{2},\frac{\beta}{2},\beta;\frac{\beta}{2}+1,\frac{\beta}{2}+1,\frac{\alpha}{2}+\beta+\frac{1}{2};1\right)+ \quad (66)$$
$${}_6F_5\left(\frac{1}{2}-\frac{\alpha}{2},\frac{\beta}{2},\frac{\beta}{2},\frac{\beta}{2},\frac{\beta}{2},\beta;\frac{\beta}{2}+1,\frac{\beta}{2}+1,\frac{\beta}{2}+1,\frac{\beta}{2}+1,\frac{\alpha}{2}+\beta+\frac{1}{2};1\right)\Big]$$

We deduce that

$$\langle X_t^4\rangle \sim 3\left(k_4(\alpha,\beta)+k_4(\beta,\alpha)\right)t^2 \quad (67)$$

### 8.3.2 Fourth cumulant

The fourth cumulant is $\kappa_4 = \langle X_t^4\rangle - 3\langle X_t^2\rangle^2 \sim 3g_4(\alpha,\beta)t^2$. Using the above, we find

$$g_4(\alpha,\beta) = \Big[k_4(\alpha,\beta)+k_4(\beta,\alpha)-\left(k_2(\alpha,\beta)+k_2(\beta,\alpha)\right)^2\Big] \quad (68)$$

No significant simplification arises in the case $\alpha=\beta$.

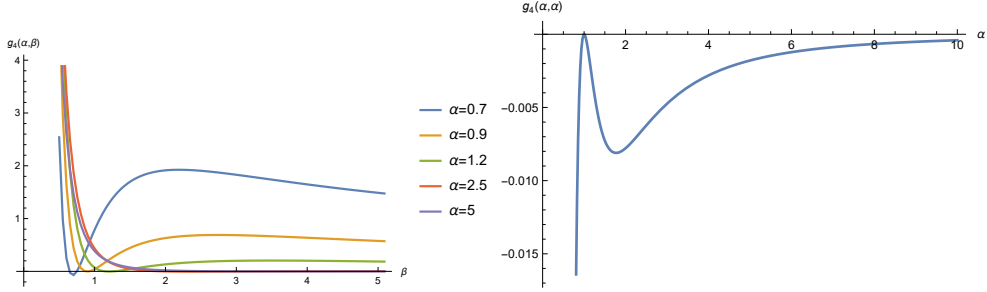

Figure 3: Left : The function $g_4(\alpha,\beta)$ plotted for different values of $\alpha$. It is interesting to note that it vanishes for nontrivial values of $\beta$. Right : the symmetric case $g_4(\alpha,\alpha)$.

## 9 Non-monotonicity of $p_{\alpha,\beta}(u)$

A non-rigorous criterion that we derived in the main text for the existence of local maxima at nonzero values of $u$ (bumps) in $p_{\alpha,\beta}(u)$ is the following. If there are bumps in the scaling function $p_{\alpha,\beta}(u)$ in the positive half-space $u>0$, then necessarily the following inequality is satisfied

$$\frac{(1-\alpha)(1-\beta)(\beta+2)^3}{\beta^2(\beta+3)(\alpha+2\beta+1)} > 1 \quad (69)$$

In particular, (69) implies that $\alpha,\beta<1$, but it is more restrictive. Note that as soon as one of $\alpha,\beta\geq 1$, there are never any bumps, so the criterion is always valid for such values. We now test whether the criterion is sufficient (ie, implies the existence of bumps) for several values of $\alpha,\beta$. The following array must be read like so : in each cell, corresponding to a couple of values $(\alpha,\beta)$, we put two symbols $A,B$. If there truly are bumps in the halfspace $u>0$ in the scaling function for these

12

| β\α | 0.2 | 0.4 | 0.6 | 0.7 | 0.8 |
|---|---|---|---|---|---|
| 0.2 | ✓,✓ | ✓,✓ | ✓,✓ | ✓,✓ | ✓,✓ |
| 0.4 | ✓,✓ | ✓,✓ | ✓,✓ | ✓,✓ | ✓,✓ |
| 0.6 | ✓,✓ | ✓,✓ | ✓, × | ✓,× | ✓,× |
| 0.7 | ✓,✓ | ✓,✓ | ✓,× | ✓,× | ×, × |
| 0.8 | ✓,✓ | ✓,✓ | ✓,× | ×, × | ×, × |

values of $(\alpha, \beta)$, we write $A = \checkmark$. If there are none, $A = \times$. The second symbol $B$ corresponds to whether or not (69) is satisfied.

As we can see, there is a relatively small range of values of $\alpha, \beta$ for which (69) fails but bumps exist nonetheless.

## 10 Regularity of the scaling function $p_{\alpha,\beta}(u)$

### 10.1 Vanishing derivative at $x = 0$

In the half-space $u > 0$, the derivative of the scaling function writes

$$\partial_u p_{\alpha,\beta}(u) \propto u \sum_{n=0}^{\infty} \frac{\left(\frac{1-\alpha}{2}\right)_n (\beta)_n}{\left(\frac{1+2\beta+\alpha}{2}\right)_n} \frac{(2n+\beta)^3}{(n+\frac{\beta-1}{2})(n+\frac{\beta+1}{2})} \frac{e^{-u^2(2n+\beta)^2}}{n!} \tag{70}$$

The order of magnitude of the summand is, as $n \to \infty$

$$\frac{\left(\frac{1-\alpha}{2}\right)_n (\beta)_n}{\left(\frac{1+2\beta+\alpha}{2}\right)_n} \frac{(2n+\beta)^3}{(n+\frac{\beta-1}{2})(n+\frac{\beta+1}{2})} \frac{e^{-u^2(2n+\beta)^2}}{n!} = O\left(n^{-\alpha} e^{-u^2(2n+\beta)^2}\right) \tag{71}$$

with $A$ a constant. Only terms in the sum with an index $n$ satisfying $u^2(2n+\beta)^2 \lesssim 1$, ie $n \lesssim 1/u$ will contribute to the sum. Therefore, (70) vanishes in the limit $u \to 0$ if and only if $u \sum_{n=1}^{1/u} n^{-\alpha} \underset{u \to 0}{\to} 0$.

It is the case, as $u \sum_{n=1}^{1/u} n^{-\alpha} = u \times O(\frac{1}{u^{1-\alpha}}) = O(u^\alpha) \to 0$. Hence, $p_{\alpha,\beta}$ has a vanishing right derivative. By reversing space, we conclude that it also has a vanishing left derivative. Thus, the derivative at 0 is well-defined and $\partial_u p_{\alpha,\beta}(u)_{|u=0} = 0$. Note that an extension of the same argument tells us that $p_{\alpha,\beta}$ does *not* have a second derivative at $u = 0$ as long as one of $\alpha, \beta < 1$.

### 10.2 Continuity at $u = 0$

Proving directly continuity of $p_{\alpha,\beta}$ at $u = 0$, even if this must hold physically, is complicated from the explicit expression of $p_{\alpha,\beta}$. This in fact yields an interesting, non-trivial identity. Indeed, consider the right limit, which Mathematica can compute and which writes

$$p_{\alpha,\beta}(0^+) = \beta \left( \frac{2\alpha}{\sqrt{\pi}(\alpha+\beta)} - \frac{2^{-\alpha}(\alpha-1)(\beta-1)\Gamma\left(\frac{\beta+1}{2}\right)\Gamma(\alpha+\beta)\,{}_3\tilde{F}_2\left(\frac{3-\alpha}{2}, \frac{\beta+1}{2}, \beta+1; \frac{\beta+5}{2}, \frac{\alpha+3}{2}+\beta; 1\right)}{\Gamma\left(\frac{\alpha}{2}\right)} \right) \tag{72}$$

13

The left limit is deduced from swapping $\alpha, \beta$. The equality of both limits, which is equivalent to the continuity of $p_{\alpha,\beta}$, implies the following identity of hypergeometric functions

$$h(\alpha, \beta) = \Gamma(\beta+1) \frac{{}_3F_2\left(\frac{3-\alpha}{2}, \frac{\beta+1}{2}, \beta+1; \frac{\beta+5}{2}, \frac{\alpha+3}{2}+\beta; 1\right)}{\Gamma(\frac{\beta+5}{2})\Gamma(\frac{\alpha+3+2\beta}{2})} = h(\beta, \alpha) \tag{73}$$

Mathematica is not able to prove the above formally. However, we can see numerically that $h(\alpha, \beta) - h(\beta, \alpha) = 0$ with precision up to $1e-16$.



\end{document}